\documentclass[runningheads]{llncs}
\usepackage[T1]{fontenc}
\usepackage{graphicx}
\usepackage{hyperref}
\usepackage{color}

\usepackage{algcompatible}
\usepackage{algorithm}

\usepackage{rotating}

\usepackage{amsmath}
\usepackage{subcaption}

\usepackage{soul}
\usepackage{tikz}
\usetikzlibrary{calc}
\makeatletter
\newif\if@anonymize

\@anonymizefalse  

\if@anonymize
\newcommand{\highlight@DoHighlight}{
	\fill [outer sep = -15pt, inner sep = 0pt, color=black]
	($(begin highlight)+(0,8pt)$) rectangle ($(end highlight)+(0,-3pt)$) ;
}

\newcommand{\highlight@BeginHighlight}{
	\coordinate (begin highlight) at (0,0) ;
}

\newcommand{\highlight@EndHighlight}{
	\coordinate (end highlight) at (0,0) ;
}

\newdimen\highlight@previous
\newdimen\highlight@current
\newlength{\item@width}

\DeclareRobustCommand*\anonymize{%
	\SOUL@setup
	\def\SOUL@preamble{%
		\begin{tikzpicture}[overlay, remember picture]
			\highlight@BeginHighlight
			\highlight@EndHighlight
		\end{tikzpicture}%
	}%
	\def\SOUL@postamble{%
		\begin{tikzpicture}[overlay, remember picture]
			\highlight@EndHighlight
			\highlight@DoHighlight
		\end{tikzpicture}%
	}%
	\def\SOUL@everyhyphen{%
		\discretionary{%
			\SOUL@setkern\SOUL@hyphkern
			\SOUL@sethyphenchar
			\tikz[overlay, remember picture] \highlight@EndHighlight ;%
		}{%
		}{%
			\SOUL@setkern\SOUL@charkern
		}%
	}%
	\def\SOUL@everyexhyphen##1{%
		\SOUL@setkern\SOUL@hyphkern
		\settowidth{\item@width}{##1}%
		\makebox[\item@width]{}%
		\discretionary{%
			\tikz[overlay, remember picture] \highlight@EndHighlight ;%
		}{%
		}{%
			\SOUL@setkern\SOUL@charkern
		}%
	}%
	\def\SOUL@everysyllable{%
		\begin{tikzpicture}[overlay, remember picture]
			\path let \p0 = (begin highlight), \p1 = (0,0) in \pgfextra
			\global\highlight@previous=\y0
			\global\highlight@current =\y1
			\endpgfextra (0,0) ;
			\ifdim\highlight@current < \highlight@previous
			\highlight@DoHighlight
			\highlight@BeginHighlight
			\fi
		\end{tikzpicture}%
		\settowidth{\item@width}{\the\SOUL@syllable}%
		\makebox[\item@width]{}%
		\tikz[overlay, remember picture] \highlight@EndHighlight ;%
	}%
	\SOUL@
}
\else
\newcommand{\anonymize}[1]{#1}
\fi
\makeatother

\begin{document}
\title{Skeletal Data Matching and Merging from Multiple RGB-D Sensors for Room-Scale Human Behaviour Tracking}
%
\titlerunning{Skeletal Data Matching and Merging from Multiple RGB-D Sensors}
%
\author{Adrien Coppens\inst{1}\orcidID{0000-0002-2841-6708} 
\and Valerie Maquil\inst{1}\orcidID{0000-0002-0198-3729}
}
%
\authorrunning{\anonymize{A. Coppens et al.}}
%
\institute{Luxembourg Institute of Science and Technology, Esch-sur-Alzette, Luxembourg
\email{firstname.lastname@list.lu}}
%
\maketitle              
\begin{abstract}

A popular and affordable option to provide room-scale human behaviour tracking is to rely on commodity RGB-D sensors 
as such devices offer body tracking capabilities at a reasonable price point.
While their capabilities may be sufficient for applications such as entertainment systems where a person plays in front of a television, RGB-D sensors are sensitive to occlusions from objects or other persons that might be in the way in more complex room-scale setups.
To alleviate the occlusion issue but also in order to extend the tracking range and strengthen its accuracy, it is possible to rely on multiple RGB-D sensors and perform data fusion. Unfortunately, fusing the data in a meaningful manner raises additional challenges related to the calibration of the sensors relative to each other to provide a common frame of reference, but also regarding skeleton matching and merging when actually combining the data.
In this paper, we discuss our approach to tackle these challenges and present the results we achieved, through aligned point clouds and combined skeleton lists. These results successfully enable unobtrusive and occlusion-resilient human behaviour tracking at room scale, that may be used as input for interactive applications as well as (possibly remote) collaborative systems.


\end{abstract}
\section{Introduction}



Room-scale interaction tracking concerns many scenarios and setups, with varying numbers of users and interaction needs beyond those of standard desktops. 
Various types of interactive surfaces including displays might be spread across the room.
As part of an ongoing project, we work with wall-sized displays, which are frequently used to present large amounts of data and to support collaborative work and decision-making in group settings. The peculiarity of that project is that we look at interconnecting two such displays to enable remote collaboration scenarios. As the resulting interactions happen at room-scale, we need a human behaviour tracking and analysis system adapted to such physical space.

Wall-sized displays are often equipped with touch frames that provide natural and efficient interaction for users standing close to the display~\cite{oehl2007considerations}. However, touch input is not always sufficient, both in terms of interaction options (as they involve physical navigation~\cite{jakobsen2012proximity}) and with regards to behaviour tracking, e.g. as input to support our remote collaboration system.

Many existing tracking systems rely on hand-held or body-worn equipment to enable distant interaction capabilities, e.g. through telepointers~\cite{kopper2008increasing,febretti2013cave2}, markers~\cite{vogel2005distant} or full-on body suits~\cite{kim2018deep}.
While these options typically provide good tracking accuracy and demonstrate high reliability and robustness to occlusions~\cite{nancel2011mid,nancel2013high}, they can be obtrusive as they require users to hold or wear additional devices.

Therefore, previous work has explored the use of unobtrusive tracking devices such as video cameras.
By processing the recordings of such standard (RGB) cameras, it is possible to enable skeletal tracking (e.g. through OpenPose~\cite{openpose}, that tracks whole bodies, including hands and faces specifically~\cite{openpose-hand-face}).

That approach might be sufficient for some cases but the issue with these RGB cameras is that they are limited to two-dimensional input images, which makes three-dimensional inferences complex. While some have tackled that issue by using several RGB cameras~\cite{chu2008real}, many researchers have instead opted for RGB-D sensors~\cite{camplani2017multiple}, that also gather depth information (e.g. through Time-of-Flight~\cite{kolb2010time} measurements). Such devices enable better skeletal tracking capabilities to an extent that enables gaze~\cite{bai2020user}, gesture~\cite{kim2015depth} and mid-air interactions~\cite{polacek2012comparative}.

However the capture range of RGB-D sensors is limited and cannot cover the whole space required for room-scale tracking. Furthermore, when several objects or users are present, occlusion problems might occur frequently. 

In this paper, we study the feasibility of relying on multiple RGB-D sensors to support room-scale tracking. The combination of the resulting data raises additional challenges, including the gathering of knowledge on the relative placement of the sensors, but also the skeletal data fusion itself. To tackle these challenges, we describe our approach consisting of i) multi-sensor calibration, ii) skeleton matching, and iii) skeleton merging, and present the rationale behind the implementation in each of these steps. We then discuss the potential merits and limitations of that system.

%


\section{Multi-sensor calibration}

In order to feed our room-scale tracking system, we rely on two Azure Kinect RGB-D sensors, that each continuously provide depth images as well as body tracking data (through the Azure Kinect Sensor SDK\footnote{\url{https://github.com/microsoft/Azure-Kinect-Sensor-SDK}}).
To facilitate the synchronisation and combination of the input data from these sensors, we rely on the \textbackslash psi framework~\cite{bohus2021platform} to build our processing pipeline. The framework comes with components ensuring the easy integration of Azure Kinect devices, but also includes data manipulation, recording and visualisation tools, which helps in alleviating some of the burden of dealing with streams of 
temporal data.

For example, simultaneous captures of the same tracking area by multiple sensors would result in significant inaccuracies. The hardware sync feature of the SDK\footnote{\url{https://learn.microsoft.com/en-us/azure/kinect-dk/multi-camera-sync}} solves that issue by making sure the capture intervals are shifted, by relying on connections through 3.5-mm audio cables between the sensors. These time modulations introduce offsets between captures from different but overlapping sensors, but the resulting adversary effects can easily be reduced through time-based interpolation with \textbackslash psi.


Solving this time synchronisation issue is however insufficient to combine data from multiple sensors as part of a room-scale tracking system, as their relative placement (to each other or to a reference coordinate system) still needs to be known, so that the data they gather separately may be converted into a common frame of reference.
In the literature, this is typically represented by a transformation matrix (one per sensor) that contains what is often referred to as \textit{extrinsic} (or sometimes external) parameters. That \textit{extrinsic matrix} essentially represents the sensor's pose in the \textit{world coordinate system}, and enables transformations between the sensor and that world~\cite{zhang2021camera}.

For a well-defined setup with precise measurements regarding the dimensions and placement of the RGB-D sensors, it is of course possible to opt for a manual approach in which the data is provided directly to the system by a human. 

A common approach to avoid potentially tricky placements and measurements is to rely on markers and known 2D motifs for calibration, in particular based on checkerboards~\cite{maimone2011encumbrance} whose square-based patterns are recognised by two sensors to determine the extrinsic parameters. The previously mentioned Azure Kinect SDK even provides an example project using that approach\footnote{\url{https://github.com/microsoft/Azure-Kinect-Sensor-SDK/tree/develop/examples/green_screen}}. 

We therefore tried the checkerboard approach based on the aforementioned example project but the results we obtained were not satisfactory. This might partly be due to our reliance on a checkerboard we printed ourselves on an A3 sheet of paper that we placed on a sturdy cardboard plate, as a ``commercial'' checkerboard that might have produced better calibration outcomes.

Another possibility is to capitalise on the tracking data itself: with only one person in the tracking area being tracked by the two sensors, the corresponding skeletons (one per sensor) can be used as a reference~\cite{baek2015dancea}.


We tried that approach, by selecting a set of candidate joints
to act as references from which we compute an average transformation matrix. We do so by splitting translations (straightforward average computation) and rotations transformations to average them separately (see Hartley et al.~\cite{hartley2013rotation} for a tutorial on rotation averaging). Note that only candidate joints whose reported tracking state was at least ``medium'' were used in the computation. While the calibration process with that approach was convenient, the results were unsatisfactory and led to ambiguities with the skeleton matching logic we describe in section~\ref{sec:matching}.

\begin{figure}
    \centering
    \includegraphics[width=.9\linewidth]{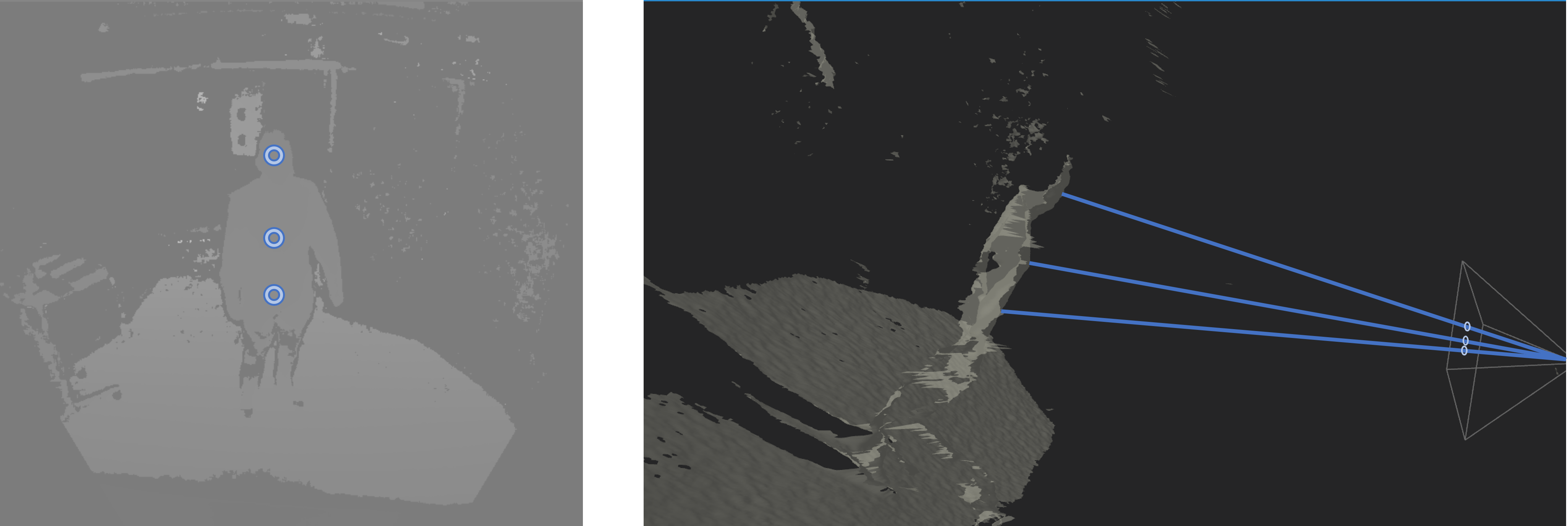}
    \caption{Generation of a partial point cloud (right) from a depth image (left). The blue circles show a selection of three points on the depth image, replicated on the image plane seen on the right picture. As these points have a similar colour on the depth image, they are at similar distance from the sensor (visualised as blue lines of similar length on the right picture) and the combination of that distance information with a projection of the image points using the sensor's intrinsic parameters generates resulting points to be included in the point cloud.}
    \label{fig:pointcloud-from-depth}
\end{figure}

Another opportunity provided by RGB-D sensors is that they generate depth data that may be converted point clouds, as illustrated in Figure~\ref{fig:pointcloud-from-depth}.
As seen in the Figure, the generated point cloud can be considered as partial in the sense that it can only ever contain the front facing part of any entity in range (any possible line starting from the sensor may only lead to a maximum of one corresponding point that belongs to the closest element encountered along that line).

Two sensors roughly looking at the same scene should generate similar point clouds, which can therefore be matched to compute the transformation matrix between the source sensors.
A popular (family of) algorithm(s) to perform such matchings is named Iterative Closest Point (ICP)~\cite{wang2017survey}. 

We therefore decided to implement an approach based on that algorithm. As made clear by its name, ICP is an iterative process, that requires an initialisation step to be used as a starting point. We chose to simply use the skeleton-based calibration described above as that starting point.



However, the initial results we obtained with the ICP approach were also insufficient, in the sense that the floor looked aligned but objects on it were not
. 
This was because planar surfaces such as floor parts will generate very similar (planar) point clouds. Assuming the floor is homogeneously leveled in the target room, running ICP on point clouds that contain many floor points will likely result in a combined floor that is indeed aligned.
The problem is that our setups rely on pairs of sensors that are capturing their respective scene from different viewpoints
and that therefore do not contain that many points from commonly seen surfaces. This means that the floor points largely outnumber the other surfaces and ICP will essentially perform the alignment mostly based on floor points that are mostly not even from the same exact portions of the floor.

We therefore decided to filter the generated point clouds in an attempt to focus on points that should mostly correspond to the same surfaces (and should therefore indeed be matched by ICP to generate a good transformation matrix).

\begin{figure}
\centering
\begin{subfigure}{0.243\textwidth}
    \includegraphics[height=4.3cm]{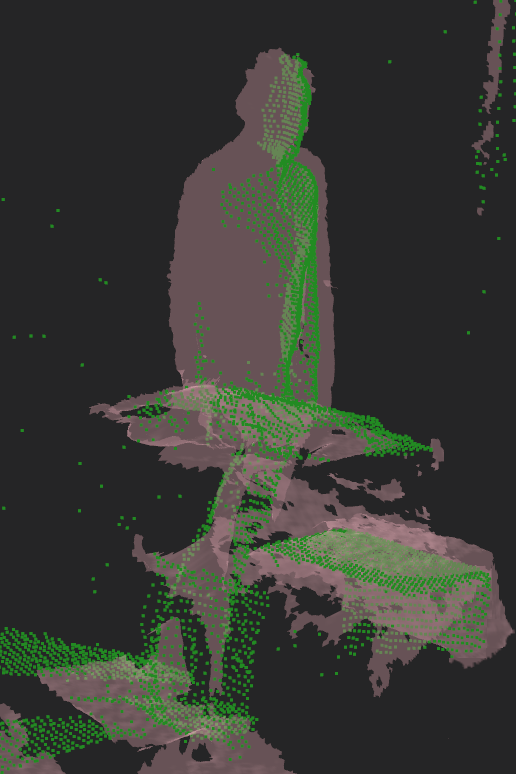}
    \caption{Single user \, (front-right view).}
    \label{fig:pc-icp-filtered-frontright}
\end{subfigure}
\begin{subfigure}{0.443\textwidth}
    \includegraphics[height=4.3cm]{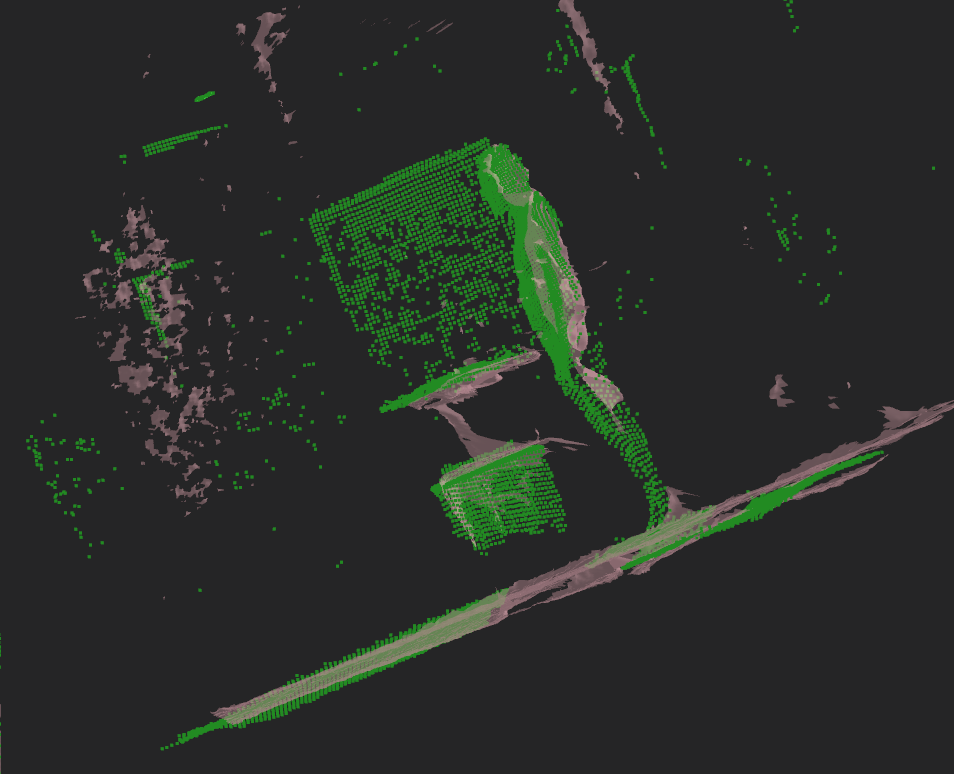}
    \caption{The same user from the same \, \mbox{capture}, from a different (left) view.}
    \label{fig:pc-icp-filtered-left}
\end{subfigure}
\begin{subfigure}{0.298\textwidth}
    \includegraphics[height=4.3cm]{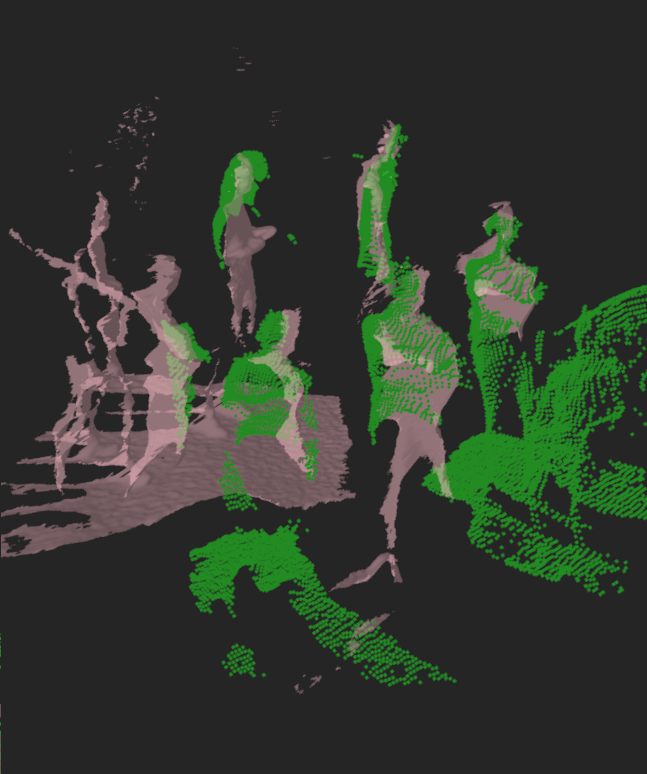}
    \caption{Another capture with users scattered in a room.}
    \label{fig:pc-icp-filtered-many}
\end{subfigure}
        
\caption{Calibration results using the ICP approach with filtered point clouds, with green and red (full) point clouds corresponding to two separate sensors. In the first two subfigures, the shapes of calibration objects (a standing desk and a cardboard box) may be perceived, with the single user standing near them.}
\label{fig:pc-icp}
\end{figure}

To do so, we place a few non-reflective objects (cardboard box, chair, etc.) in both of their tracking ranges. we place such objects in line, about 2.5-3 meters away from both sensors, to maximise the jointly tracked surfaces while staying at good tracking distance.
We then ask the person (whose skeleton is used as common reference to initiate the ICP process) to stand close to these objects and only consider point cloud points that are close to that person
. This normally result in two point clouds that contain many common points for the floor below the reference person, but also the previously mentioned jointly tracked (``front-facing'') surfaces as well as the front part of the person's body. Our hope with that approach is that the inclusion of floor points should produce an ICP result whose transformation matrix will lead to fused data where floor points are aligned. 
The jointly tracked surfaces should themselves help with the rest of the degrees of freedom and the combination of floor and front-facing surface points should therefore lead to pairs of point clouds suitable for alignment. After a couple of ICP runs following the aforementioned process, we indeed ended up with properly aligned results, as seen in Figure \ref{fig:pc-icp}.

\section{Skeleton Matching}
\label{sec:matching}
A proper calibration is enough to generate combined point clouds that may then cover bigger areas as we may transform the data from the different sensors within the same room as being expressed in the same coordinate system.

\begin{figure}
\centering
\begin{subfigure}{0.47\textwidth}
    \includegraphics[width=.9\textwidth]{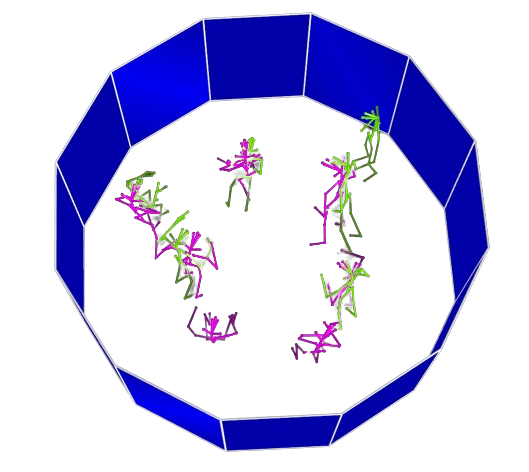}
    \caption{Skeletons from two (green- and pink-coded) sensors. Notice the isolated skeletons: a green one on the top right, and two pink ones in the bottom.}
    \label{fig:skeletonmatching-twosets}
\end{subfigure}
\hfill
\begin{subfigure}{0.49\textwidth}
    \includegraphics[width=.9\textwidth]{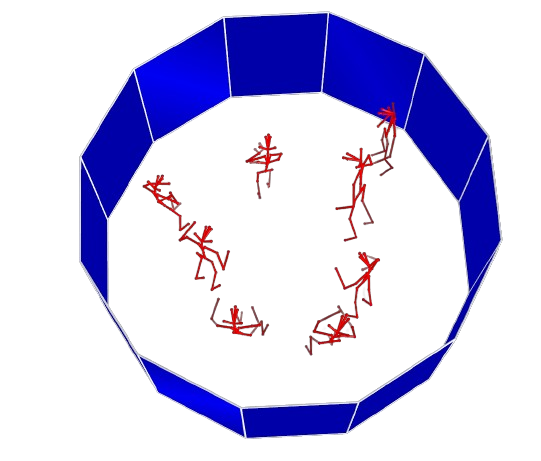}
    \caption{The merged skeletons following our skeletal data fusion algorithm, showing the successful pairing of overlapping skeletons but also the inclusion of isolated ones.}
    \label{fig:skeletonmatching-merged}
\end{subfigure}
\caption{The skeleton matching problem, with skeletons from two different sensors that are matched to form a merged set of skeletons. While most persons in the room are tracked by both sensors and therefore result in two overlapping skeletons, some are isolated as only one of the sensors currently sees them.}
\label{fig:skeletonmatching}
\end{figure}

While we could theoretically merge the transformed point clouds and use the result as input for a hypothetical body tracking model, most of the existing tools relying on RGB-D sensors expect non-fused data, in the sense that they were trained with setups involving only one sensor and front-facing users. Providing merged point clouds to them might lead to inaccurate results, and we therefore opted to perform the data fusion at the skeletal tracking level, i.e. analyse the individual sets of skeletons coming from each sensor to decide whether (and which) skeletons from different sets belong to the same person.
This is illustrated through a rather crowded case in Figure \ref{fig:skeletonmatching}, where input skeletons from two sensors (Figure \ref{fig:skeletonmatching-twosets}) are combined into a merged set (Figure \ref{fig:skeletonmatching-merged}).

After converting the body tracking information to a common coordinate system and filtering non-suitable skeletons (too close to the sensor or outside the tracking area), our logic for dealing with skeleton matching boils down to 4 steps: 
(1) settle easy cases where it seems clear which skeletons to match based on a distance threshold,
(2) perform a first pass through the remaining ambiguities to see whether some where lifted by the previous step which might have removed all but one candidates,
(3) perform a second pass through the remaining ambiguities to see whether previous matchings help in choosing one candidate,
(4) solve the remaining ambiguities through a greedy approach that minimises the maximum inter-pelvis distance of the worst matching.
Following that 4-step process, we produce a list combining both matched and isolated skeletons. 

\section{Skeleton Merging} 

While isolated skeletons do not require any further processing, pairs of matched skeletons must be merged to form one single skeleton. Many approaches to perform this merging have been explored. They typically filter or attribute weights to data based on known or observed properties of the tracking system, such as whether a given joint is currently occluded for a given sensor.
As we primarily put our focus on the matching logic, the merging of skeletons we perform simply relies on a weighted average based on the tracking confidences reported by the Azure Kinect SDK. We associate a reported confidence of ``None'' to a weight of $0$, ``Low'' to $0.25$, ``Medium'' to $0.5$, and ``High'' to $1$, although that latter value does not seem to practically be used at all in the current version of the SDK.

For joint $i$ with position $\overrightarrow{p_i}$ and whose orientation is encoded as three axis vectors $\overrightarrow{x_i}$, $\overrightarrow{y_i}$ and $\overrightarrow{z_i}$, we therefore define a weight $w_i$ based on the above confidence associations. 
Similarly, we define a weight $w_j$ for joint $j$, with position $\overrightarrow{p_j}$ and orientation $(\overrightarrow{x_j}, \overrightarrow{y_j}, \overrightarrow{z_j})$.
We compute the merged position $\overrightarrow{p_m} = \overrightarrow{p_i} + \frac{w_j}{w_i + w_j} * (\overrightarrow{p_j} - \overrightarrow{p_i})$, as well as the merged orientation $(\overrightarrow{x_m}, \overrightarrow{y_m}, \overrightarrow{z_m})$ through: 
\begin{equation}
    \begin{cases}
      \overrightarrow{x_m} = & \overrightarrow{x_i} + \frac{w_j}{w_i + w_j} * (\overrightarrow{x_j} - \overrightarrow{x_i})\\
      \overrightarrow{y_m} = & \overrightarrow{y_i} + \frac{w_j}{w_i + w_j} * (\overrightarrow{y_j} - \overrightarrow{y_i})\\
      \overrightarrow{z_m} = & \overrightarrow{z_i} + \frac{w_j}{w_i + w_j} * (\overrightarrow{z_j} - \overrightarrow{z_i})
    \end{cases}       
\end{equation}

\section{Discussion}
\label{sec:discussion}

The presented work essentially describes the implementation process and the rationale behind our multi-sensor system for room-scale behaviour tracking, based on the literature and informal tests with the project team. 

One of the main challenges was to obtain a suitable transformation matrix to convert the tracking data to a common coordinate system. Based on our experiences, we found that this can be best tackled with an ICP-based approach. Regarding the skeletal matching, tracking inaccuracies led to difficulties in pairing skeletons based on proximity alone and there is a need to lift ambiguities in a meaningful manner. Finally, the skeletal merging method was more straightforward, as the weighted average approach we opted for seemed to produce satisfactory results.

By fusing such data from separate RGB-D sensors, we extend the tracking range of our system and make it more robust to occlusions, which effectively strengthen the system's tracking stability and accuracy for jointly tracked areas. This tracking system and the resulting implementation show great potential as a basis to be built upon, for behaviour and collaboration analysis but also for distant interaction. As we aim at enabling efficient and pleasant remote collaboration across distant wall-sized displays, the skeletal data gathered by this tracking system will be extremely valuable to understand and transmit the necessary collaboration information to the remote side.

\label{sec:future}


A natural extension to the presented work will be to add an extra scene calibration step, in which we will gather knowledge on the placement of interactive surfaces relative to the coordinate system of the tracking system. This will allow to link tracking data such as pointing and gazing information to target location on these surfaces, including interactive displays.

Our matching logic could also be made more ``defensive'', through approaches such as delays before a matching is created or removed, or before isolated skeletons are taken into account. We could also simply ignore isolated skeletons until they are matched confidently (e.g. with strict proximity conditions for all joints).
These approaches would incur a cost related to the rapid and exhaustive inclusion of users, but would likely improve the stability and reliability of the matchings.


As part of our future work, we plan on implementing and evaluating these alternatives, but also explore additional options regarding the merging strategy, through formal evaluations based on user studies.


 


\section{Limitations}
\label{sec:limitations}

Despite the addition of a second sensor, some occlusions can still happen, be it with two users hiding a third one (in between them) from the two sensors, or because of a ``tracking dead zone'' where none of the two sensors can capture data from.
More sensors could be added following the same calibration and data fusion procedure but this would increase both the needs in terms of computing power and the risks of some of the skeletons being wrongly matched.

Another limitation of our setup is the reliance on Azure Kinect sensors which have been retired as of 2023, although the underlying technology has been transferred and alternatives do exist
. This is however simply a limitation of our practical setup as the procedures, algorithms, and general logic presented in the paper remain valid with other RGB-D sensors.


\section{Conclusion}
\label{sec:conclusion}

In this paper, we contributed a multi-sensor tracking system for unobtrusive room-scale 
human behaviour analysis based on affordable RGB-D sensors. By presenting our rationale and the challenges we tackled while implementing it, we contributed to the knowledge on building such systems.
This covers approaches 
for determining the placement of the sensors themselves but also tracking data fusion techniques. 
We then discussed the proposed system and the remaining limitations. 

In its current state, the implemented system is suitable for room-scale tracking and may readily provide good input to run behavioural analysis on users. It will be extended to drive distant interactions with interactive surfaces and displays and will be the foundation on which we built an awareness support system for remote collaboration across wall-sized displays.



\begin{credits}
\subsubsection{\ackname} Authors would like to thank \anonymize{the Luxembourg National Research Fund (FNR) for funding this research work under the FNR CORE ReSurf project (Grant nr C21/IS/15883550)}.

\subsubsection{\discintname}
The authors have no competing interests to declare that are
relevant to the content of this article.
\end{credits}
%
%
%
\bibliographystyle{splncs04}





\end{document}